\documentclass[preprint,amsmath,amssymb,showpacs]{revtex4}
\usepackage{graphicx}
\begin{document}
\title{Self-avoiding walks
  on a bilayer Bethe lattice}
\author{Pablo Serra}
\email{serra@famaf.unc.edu.ar}
\affiliation{Facultad de Matem\'atica, Astronom\'{\i}a y F\'{\i}sica,
Universidad Nacional de C\'ordoba, C\'ordoba, Argentina and IFEG-CONICET,
Ciudad Universitaria,
X5016LAE C\'ordoba, Argentina}
\author{J\"urgen F. Stilck}
\email{jstilck@if.uff.br}
\affiliation{Instituto de F\'{\i}sica and
National Institute of Science and
Technology for Complex Systems
\\
Universidade Federal Fluminense\\
Av. Litor\^anea s/n\\
24210-346 - Niter\'oi, RJ\\
Brazil}
\date{\today}

\begin{abstract}
We propose and study a model of polymer chains in a bilayer. Each chain 
is confined in one of the layers and polymer bonds on first neighbor
edges in different layers interact. We also define and comment results for
a model with interactions between monomers on first neighbor sites of 
different layers. The thermodynamic properties
of the model are studied in the grand-canonical formalism and both
layers are considered to be Cayley trees. In the core region of the 
trees, which we may call a bilayer Bethe lattice, we find a very rich 
phase diagram in the parameter space defined by the two activities of
monomers and the Boltzmann factor associated to the interlayer 
interaction between bonds or monomers.
Beside critical and coexistence surfaces, there are tricritical,
bicritical and critical endpoint lines, as well as higher order 
multicritical points. 
\end{abstract}


\maketitle

\section{Introduction}
\label{intro}
The theoretical study of the thermodynamic behavior of polymers, 
both in a melt or in solution,
has a long history, continuous and lattice models were proposed to
study such systems \cite{f66}. In lattice models, the first approach
is to consider the linear polymeric chains to be random walks,
this model is sometimes called {\em ideal chain
  approximation}. A more realistic approach is to represent the linear 
polymeric chains by self- and mutually avoiding walks (SAW's), and
therefore one of the simplest models is athermal, since only the
excluded volume interactions are taken into account. Regarding the
phase transitions which happen in such models, the excluded volume
interactions are essential, at space dimensions below the upper
critical dimension, to produce the correct critical exponents
\cite{dg79}. At and above the upper critical dimension the ideal chain
exponents are found. In a grand-canonical ensemble, at low activities
of monomers, a non-polymerized phase is stable, but as the activity is
increased, a transition to a polymerized phase happens. This
transition is usually continuous.

While a simple model with excluded volume interactions may explain the
polymerization transition when the chains are placed in a good
solvent, in bad solvents and at low temperatures collapsed configurations 
are favored energetically. Below the temperature ($\Theta$ point), 
at which the transition between the extended and the collapsed
configurations happens, the
polymerization transition is discontinuous, so that the collapse
transition corresponds to a tricritical point. This transition 
is frequently named coil-globule transition in the literature. A simple
effective model which explains this phenomena includes {\em
  attractive} interactions between monomers placed at first neighbor
sites which are not consecutive along a chain, these interactions 
favor more compact polymer configurations,
thus reducing the contact area of the polymer chain with the
solvent. In the literature, these interacting walks are called {\em
  self-attractive self-avoiding walks} (SASAW's) \cite{dg79}.
When it was realized by De Gennes that the simple polymerization
model could be mapped on the ferromagnetic $n$-vector model in the
limit when the number of components of the order parameter $n$
vanishes \cite{dg72}, the then new ideas of renormalization group,
in the form of and expansion of critical exponents in
$\epsilon=4-d$ with coefficients which are functions of $n$, where 
$d$ is the spacial 
dimensionality of the system, could be directly applied to
models for polymers. A detailed discussion of this limit may be 
found in \cite{wp81}. Later, the relation between polymers and magnetic
models was extended to SASAW's \cite{dg75}. Related to this problem of the 
collapse transition of a polymer chain with effective attractive interactions,
a related problem of two chemically distinct chains with attractive 
interactions between them has also been studied in the literature. As
in the case of SASAW's, the phase diagrams of such systems also display
a tricritical point, the collapsed state corresponds in this case to a 
zipped state, where steps for which the bonds of both chains move side
by side are favored. The solution of this model on fractal lattices such as 
the $2D$ Sierpinski gasket \cite{ks93} and truncated $n$-simplex lattice 
\cite{ks95}, where real space renormalization transformations are exact, 
leads to phase diagrams with tricritical points also, as is the case for 
SASAW's.

Although it is rather natural to consider the effective attractive 
interactions in
the SASAW's to be between monomers, as described above, in the literature
an alternative model, with interactions between bonds in the same 
elementary polygon of the lattice (plaquette) has also received much 
attention, one of the reasons for this is that this model may be
mapped on the $n$-vector model with four spin interactions \cite{bn89,wp81b}.
One may imagine that the two SASAW's models, with interactions between
monomers or bonds, should lead to similar thermodynamic properties, but
at least on two-dimensional lattices this is not the case. Qualitatively
different phase diagrams were found for both models on a four 
coordinated Husimi lattice
\cite{sms96} and in transfer matrix calculations combined with FSS 
extrapolations on the square lattice \cite{mos01}.
 
The behavior of magnetic models on bilayers and multilayers has
attracted attention in the literature for quite some time. For example, 
coupling through non magnetic layers shows
oscillatory behavior and may lead to giant magnetoresistance effects
\cite{b88}. The phase diagrams of magnetically disordered bilayers
have been studied on a system of two adjacent coupled Bethe lattices,
for competing intralayer interaction parameters, the phase diagram
displays multiple reentrant behavior \cite{ld92}. A variety of other
magnetic systems has also been studied on similar lattices
\cite{albayrak}. Although, as expected, Bethe lattice solutions 
overestimate the region of stability of ordered phases, they may
furnish the general features of phase diagrams. 

Polymeric chains close to interfaces between two immiscible liquids
have also been studied, and it was found in a continuum model that the
chains are attracted to the interface \cite{hp86}, the maximum of the
distribution of monomers being located close to the interface, inside
the better solvent. A similar model was also studied in \cite{w86},
with a different parametrization and several mean values and
probabilities related to the conformation of the polymer chain are
evaluated. The behavior of a copolymer close to a bilayer has been
considered also in the literature, motivated by the relation of this
problem with the relevant biological problem of the localization of a
protein in a lipid bilayer \cite{ermoshkin02}. Two regimes are found
in this study regarding the localization of the chain in the
bilayer. In the first, the density of monomers is a bimodal function,
with maxima centered on the two interfaces, while the second
configuration displays a single maximum of the density in the center
of the bilayer. In the bimodal regime, few monomers are present in the
area between the interfaces and most of the chain is adsorbed on them.

\begin{figure}[ht]
\centering
\includegraphics[width=8cm]{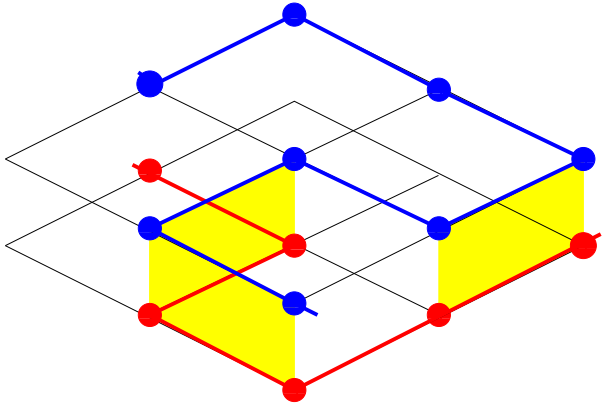}
\caption{(color online) Part of a bilayer square lattice with a polymer
chain on each layer. The three pairs of interacting bonds are edges of the
shadowed (yellow) rectangles. In the model with interactions between monomers,
five pairs of interacting monomers are present in the figure.}
\label{model}
\end{figure}

Here we address a simpler problem, with homopolymer chains only, in
the limit where each chain is entirely adsorbed on one layer and
thus no polymer bond is present linking chains adsorbed on different
interfaces. We include in the model interactions between polymer bonds
placed in different layers on corresponding edges. This interaction
takes into account the changes in statistical weight of a polymer bond
if it is in contact with the other solvent across the interface or
with a bond of another polymeric chain in the other interface. In figure 
\ref{model} a particular configuration of a part of a bilayer 
built with two square lattices is shown. Although we show
in details the calculations and results of the model with interacting
bonds, we have also studied the alternative model where the interactions
are  between monomers, and will briefly discuss the differences in 
the thermodynamic behaviors of the models. In what
follows, we consider both interfaces to be Cayley trees of arbitrary
coordination number $q$, with polymeric chains on them whose endpoints
are placed on the surface sites of the trees. Since we will study 
the behavior of the model in the core of the bilayer tree, we may 
consider this solution of the model to be a Bethe approximation for the 
model on regular lattices with the same coordination number.

The paper is organized as follows: In section \ref{defmod} we define the model 
more precisely on the bilayer Bethe lattice and obtain its solution in 
terms of fixed points of recursion relations. The fixed points which
are stable in some region of the parameter space are presented in 
section \ref{pfp}, studying the stability of
these fixed points we address the transitions between the phases
which are associated to each physical fixed point. The phase diagram in 
the three-dimensional parameter space of the model is studied in some
detail in section \ref{pd}, and final discussions and conclusion may
be found in section \ref{conc}

\section{Definition of the model and solution on the bilayer Bethe  lattice} 
\label{defmod}

\begin{figure}[ht]
\centering
\includegraphics[width=8cm]{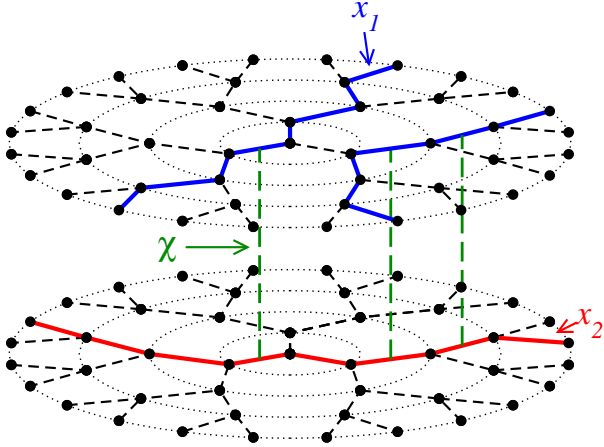}
\caption{(color online) Bilayer Bethe lattice with $q=3$ and $M=4$ generations. 
The polymer configuration has a statistical weight $x_1^{14} \,x_2^8 \, \chi^3
\,=\,x_1^{11}\,x_2^5\,\kappa^3$, where $\kappa=x_1x_2\chi$.}
\label{bbl}
\end{figure}

The problem is defined in 
the grand-canonical formalism, so that the parameters of the model are
the fugacities of a polymer bond in layer 1, $x_1=\exp[\mu_1/(k_BT)]$,
and in layer 2, $x_2=\exp[\mu_2/(k_BT)]$, as well as a Boltzmann
factor $\chi=\exp[\epsilon/(k_BT)]$ for each pair of polymer bonds of
corresponding edges or corresponding
monomers in different layers.  The energy of interaction of
these pairs of bonds or monomers is, therefore, equal to $-\epsilon$, and is
attractive if $\epsilon>0$. In what follows, we 
restrict ourselves to the model of interacting bonds, unless otherwise
stated. In figure \ref{bbl}
a particular configuration of the model is shown for a bilayer Bethe
lattice with coordination number equal to three.

The grand-canonical partition function of the model may be written as 

\begin{equation}
{\cal Y}\,=\, \sum \;x_1^{N_1} \; x_2^{N_2} \; \chi^{N_B}
\end{equation}

\noindent where the sum is over configurations of self and mutually
avoiding  walks placed on the trees. The endpoint monomers of the walks are
placed on the surface sites of the trees. $N_1$ is the number of  polymeric
bonds on the  Cayley tree 1, $N_2$ is the number of polymeric bonds on
the Cayley tree 2,  and $N_B$ is the number of pairs of polymeric
bonds located on first neighbor edges on different trees. We refer to
figure \ref{bbl} for the statistical weight of a particular
configuration of the system.

The solution of systems defined on tree-like structures defining recursive
relations for partial partition functions (ppf's) is well known
\cite{baxter}, and 
is particularly useful for polymeric models because it is simple 
to impose the excluded-volume  condition \cite{ss90,ss07}.
The generalization to a bilayer Bethe lattice is straightforward, and  was 
widely used to study magnetic bilayers \cite{ld92,albayrak}.
We thus proceed defining partial partition functions for rooted
sub-trees. The fixed root configuration of these sub-trees is
specified by the polymer bonds which arrive at the root
sites of sub-trees 1 and 2 coming from outer generations of sites. We need four
partial partition functions: $g_{0,0},\; g_{1,0},\;g_{0,1},$ and
$g_{1,1}$ where the first (second) sub-index denotes the numbers of polymers 
bonds incident at the root site of the sub-tree 1 (2). Considering the
operation of attaching $q-1$ sub-trees to a new pair of root sites, we
may write recursion relations for the partial partition functions
of a sub-tree with one additional generation of sites, denoted by primes. 
The recursion relations are: 
\begin{subequations}
\begin{eqnarray}
g'_{0,0} &=&   g_{0,0}^{q-1}\,+\,\frac{1}{2} (q-1) (q-2) 
\left(x_1^2 \,g_{1,0}^2 \,+\, x_2^2 \, g_{0,1}^2 \right)\, g_{0,0}^{q-3}\,+\, 
\nonumber  \\
\mbox{}  &\mbox{}&
(q-1) (q-2) (q-3) \, x_1^2 \, x_2^2 \,\chi \,g_{1,1} \, g_{1,0} \, g_{0,1}  \,
g_{0,0}^{q-4} \,+\, \nonumber  \\
\mbox{}  &\mbox{}&
\frac{1}{2} (q-1) (q-2) \,x_1^2 \,x_2^2 \,\chi^2 \,g_{1,1}^2 \,g_{0,0}^{q-3}
\,+\, \nonumber  \\
\mbox{}  &\mbox{}&
\frac{1}{4} (q-1) (q-2)  (q-3)  (q-4) \, x_1^2 \, x_2^2 \, g_{1,0}^2 \, 
g_{0,1}^2  \, g_{0,0}^{q-5} \,, \\
\mbox{}  \nonumber  \\
g'_{1,0}\,&=&\,(q-1) \,x_1 \,g_{1,0} \,  g_{0,0}^{q-2} \,+\,  
 (q-1) (q-2) \, x_1 \, x_2^2 \, \chi \,g_{1,1} \,  g_{0.1} \, g_{0,0}^{q-3} 
\,+\, \nonumber  \\
\mbox{}  &\mbox{}&
\frac{1}{2} (q-1) (q-2) (q-3) \, x_1 \, x_2^2 \,g_{1,0} \,  g_{0,1}^2  
\,g_{0,0}^{q-4} \,, \\
\mbox{}  \nonumber  \\
g'_{0,1}\,&=&\,(q-1) \,x_2\, g_{0,1} \, g_{0,0}^{q-2} \,+\,
(q-1) (q-2) \,x_1^2 \, x_2 \, \chi \,g_{1,1} \,  g_{1,0} \, g_{0,0}^{q-3} 
\,+\, \nonumber  \\
\mbox{}  &\mbox{}&
\frac{1}{2} (q-1) (q-2) (q-3) \,x_1^2 \, x_2\,g_{1,0}^2 \,  g_{0,1} \,
g_{0,0}^{q-4} \,,\\
\mbox{}  \nonumber  \\
g'_{1,1}\,&=&\,(q-1) \,x_1 \,  x_2 \, \chi \,g_{1,1} \, g_{0,0}^{q-2}\,+\,
(q-1) (q-2) \, x_1 \, x_2 \,g_{1,0} \,g_{0,1} \, g_{0,0}^{q-3} \,.
\end{eqnarray}
\end{subequations}

The ppf's grow exponentially with the iterations, so, as usual, it is
convenient to define ratios of the ppf's, and the thermodynamic
properties of the model may be expressed by the fixed point values of
these ratios. The ratios we use are:

\begin{equation}
R_1=x_1\,\frac{g_{1,0}}{g_{0,0}} \;,\;\;
R_2=x_2\,\frac{g_{0,1}}{g_{0,0}} \;,\;\;
R_3=x_1 \,x_2\,\chi\,\frac{g_{1,1}}{g_{0,0}} \;.
\end{equation}

The recursion relations for the ratios will be

\begin{subequations}
\label{rr}
\begin{eqnarray}
R'_1 \,&=&\,\left[(q-1) R_1 \,+\,
 (q-1) (q-2) \,R_2 R_3 \,+\,\frac{1}{2} (q-1) (q-2) (q-3) \,R_1 R_2^2\,\right] 
\frac{x_1}{D} \,,  \\
\mbox{}  &\mbox{}& \nonumber \\
R'_2\,&=&\,\left[(q-1) R_2\,+\, (q-1) (q-2) R_1 R_3\,+\, 
\frac{1}{2} (q-1) (q-2) (q-3) \, R_1^2 R_2\,\right]
\frac{x_2}{D}  \,,\\
\mbox{}  &\mbox{}& \nonumber \\
R'_3\,&=&\,\left[(q-1) R_3\,+\, (q-1) (q-2)  \,R_1 R_2\,\right] 
\frac{\kappa}{D} \,,
\end{eqnarray}
\end{subequations}

\noindent where  $\kappa \equiv x_1\,x_2\,\chi$, and

\begin{eqnarray}
D&=& 1+\frac{1}{2} (q-1) (q-2) \left( R_1^2+R_2^2+R_3^2 \right) 
+  (q-1) (q-2) (q-3) \,R_1 R_2 R_3 \,+ \nonumber \\
\mbox{}  &\mbox{}&  \nonumber \\
\mbox{}  &\mbox{}& \frac{1}{4} (q-1) (q-2) (q-3) (q-4)  R_1^2 R_2^2. 
\end{eqnarray} 

Connecting $q$ sub-trees to the central site of the Cayley tree, the
grand-canonical partition function takes the form

\begin{eqnarray}
\frac{{\cal Y}}{g_{00}^q} \,\equiv\,Y&=& 
1+\frac{1}{2} q (q-1)  \left( R_1^2+R_2^2+R_3^2 \right)
\,+\, q (q-1) (q-2) \,R_1 R_2 R_3\,+\, \nonumber \\
\mbox{}  &\mbox{}& \nonumber \\
\mbox{}  &\mbox{}&  \frac{1}{4} q (q-1) (q-2) (q-3) R_1^2 R_2^2 \,.
\end{eqnarray}

\noindent The Bethe lattice solution of the model corresponds to its
behavior in the core of the Cayley tree, so that surface effects are
eliminated. The corresponding dimensionless free energy per site may 
be found using
a prescription proposed by Gujrati \cite{g95}. The result is

\begin{equation}
\label{fe}
\phi_b\,=\,- \frac{1}{2}  \left[q \ln{(D)}\,-\,(q-2) \ln{(Y)} \right].
\end{equation}

\noindent This expression may be derived quite easily supposing that the
free energy of the model on the whole Cayley tree (divided by $k_BT$) 
may be written as $\Phi=-\ln {\cal Y}=N_s\phi_s+
N_b\phi_b$, where $\phi_s$ and $\phi_b$ are the free energies per site for the
$N_s$ surface and $N_b$ bulk sites, respectively. Considering the free energies
of the model on two Cayley trees with $M$ and $M+1$ generations in the 
thermodynamic limit $M \to \infty$ one may then obtain an expression for the
bulk free energy per site $\phi_b$. A slightly more general argument which leads
to the same result may be found in \cite{oss09}

\section{physical fixed points}
\label{pfp}

The phase diagram is obtained looking for the 
physical fixed points of the recurrence relations 
Eqs.(\ref{rr}). For physical  fixed point we understand that there exists
 a region in the thermodynamic space $(x_1,x_2,\kappa)$ where the 
fixed point is stable 
and corresponds to a global minimum of the free energy Eq.(\ref{fe}),
thus representing a thermodynamic phase. 
A fixed point $R^*  \,=\,(R_1^*,R_2^*,R_3^*)$ is stable if the Jacobian matrix
\begin{equation}
{\cal J}_{i,j} \,=\,\left. \frac{\partial R'_i}{\partial R_j} \right|_{R^*}
\end{equation}

\noindent has all the
eigenvalues less than one. We found five stable fixed points describing
five different thermodynamic phases:

\subsubsection{Nonpolymerized fixed point ($NP$)}

The $NP$ fixed point is given by  $R_1^{(NP)}=0;\,R_2^{(NP)}=0;\,R_3^{(NP)}=0$,
which corresponds to a phase where the density of polymers vanished in both 
trees. The Jacobian matrix is 

\begin{equation}
{\cal J}^{(NP)}\,=\,
\left( \begin{array}{cccc}
(q-1) \,x_1 &  0  & 0  \\ \\
0  & (q-1)\,x_2  & 0\\ \\
0 & 0 & (q-1) \,\kappa
\end{array}
\right) \;,
\end{equation}

\noindent  thus the stability limits for the $NP$ phase are given by
the conditions
\begin{equation}
x_1^{(NP)} \,=\,\frac{1}{q-1}  \;,\;\;
x_2^{(NP)} \,=\,\frac{1}{q-1}  \;,\;\;
\kappa^{(NP)}\,=\,\frac{1}{q-1}\; .
\end{equation}

\subsubsection{Fixed point with polymers on the tree 1 ($P_1$)}

The fixed point  $R_1\ne 0;\,R_2=0;\,R_3=0$, corresponds to a phase where
the density of polymers is nonzero on the Bethe lattice numbered as one, 
and zero on the other lattice. The fixed point point value of the
ratio $R_1$ is $R_1^{(P_1)}=\sqrt{2[(q-1)x_1-1]/[(q-1)(q-2)]}$, and therefore
we may obtain the elements of the Jacobian matrix in terms of the parameters
of the model. The result is

\begin{equation}
{\cal J}^{(P_1)}\,=\,
\left( \begin{array}{cccc}
\frac{2 - (q-1) \,x_1 }{(q-1) \,x_1}&  0  & 0  \\ \\
0  & \frac{[2 + (q-1) (q-3) x_1] x_2}{(q-1) \,x_1}  &
\sqrt{\frac{ 2 (q-2) \, [ (q-1)  x_1-1 ]}{q-1}}\;\frac{x_2}{x_1}\\ \\
0 & \sqrt{\frac{2 (q-2) \, [ (q-1)  x_1-1 ]}{q-1}} \frac{\kappa}{x_1}
&  \frac{\kappa}{x_1} 
\end{array}
\right) \;.
\end{equation}

\noindent  For this phase,  the stability limits are given by

\begin{equation}
\label{p1l}
x_1^{(P_1)} \,=\,\frac{1}{q-1}  \;,\;\;\
\kappa\,=\,x_1\frac{2 x_2 + (q-1) x_1 [(q-3) x_2-1]}
{(q-1) [ 2 x_2 - (q-1) x_1 x_2-x_1]}\;.
\end{equation}

\subsubsection{Fixed point with polymers on the tree 2 ($P_2$)} 

The fixed point  $R_1= 0;\,R_2\ne 0;\,R_3=0$ corresponds to a phase
 where
the density of polymers is nonzero in the Bethe lattice numbered as two,
and vanishing on the other lattice.
By symmetry, the stability limits of the $P_2$  fixed point  
may be obtained from the stability limits of the $P_1$ fixed point
Eq. (\ref{p1l}) interchanging $x_1$ and $x_2$.

\subsubsection{Polymerized fixed point ($P$)}

  For this fixed point $R_1\ne 0;\,R_2\ne 0;\,R_3\ne 0$, so that
 the density of polymers is nonzero on both trees. In this case the 
stability lines were calculated numerically.

\subsubsection{Bilayer fixed point ($P_B$)}

This fixed point $R_1= 0;\,R_2= 0;\,R_3\ne 0$
corresponds to a phase where all polymer bonds are placed in pairs on
corresponding edges of both trees, so that the polymer of both layers
are fully correlated. The Jacobian matrix of this fixed point is

\begin{equation}
{\cal J}^{(P_B)}\,=\,
\left( \begin{array}{cccc}
\frac{x_1}{\kappa} &  \sqrt{\frac{2 (q-2) ((q-1) \kappa-1)}{q-1}} \,
\frac{x_1}{\kappa}   & 0  \\ \\
\sqrt{\frac{2 (q-2) ((q-1) \kappa-1)}{q-1}}\, \frac{x_2}{\kappa} &
\frac{x_2}{\kappa}  & 0\\ \\
0 & 0 & \frac{2-(q-1) \,\kappa}{(q-1) \,\kappa}
\end{array}
\right) \;,
\end{equation}

\noindent and the stability limits are given by

\begin{equation}
\label{pbl}
x_2^{(P_B)}\,=\,\frac{(q-1) (\kappa - x_1) \kappa}{ 
(q-1) (1 + 2 (q-2) x_1) \kappa- (3 q-5) x_1}   \;,\;\; 
\kappa^{(P_B)}\,=\,\frac{1}{(q-1)}\;.
\end{equation}

\section{phase diagram}
\label{pd}

The next step in order to obtain the phase diagram is to study the overlap
between stability regions of different physical fixed points.
If two regions where a unique fixed point is stable are separated by a common
stability line, on this line a continuous phase transition is
expected. If two or more fixed points are stable in the same region of
the parameter space, this signals coexistence of phases. In this case,
the loci of the discontinuous transitions may be found requiring that
the coexisting phases should have the same (minimum) free energy.
The five fixed points are present for coordination numbers
$q \ge 3$, but for $q=3$ the equations are simpler and thus some of the
calculations may be done analytically. Therefore, we performed all
calculations below for $q=3$. Some changes in the phase diagrams may
arise for $q>3$.

\begin{figure}[ht]
\centering
\includegraphics[width=8cm]{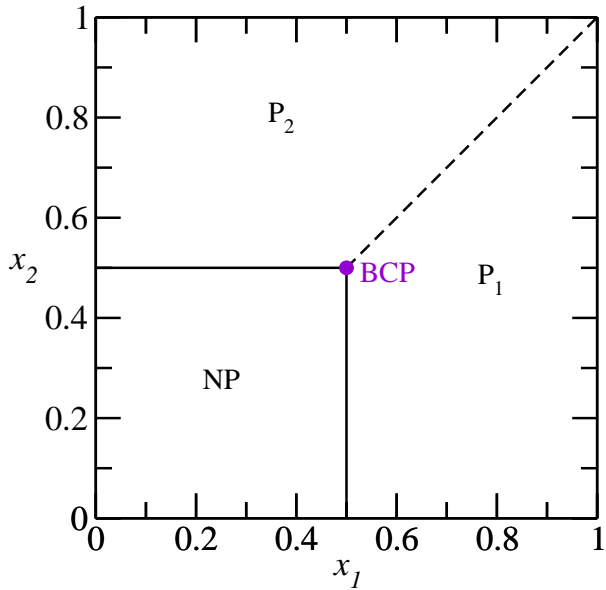}
\caption{(Color online) Phase diagram for $\kappa=0$. The dashed
  (black) curve is a first order transition and the full (black)
  lines are continuous transitions between the $NP-P_1$ and $NP-P_2$
  phases. The bicritical point BCP is represented by a (violet) dot.
   This and the following diagrams were all obtained for $q=3$.}
\label{q3k0}
\end{figure}

\begin{figure}[ht]
\centering
\includegraphics[width=8cm]{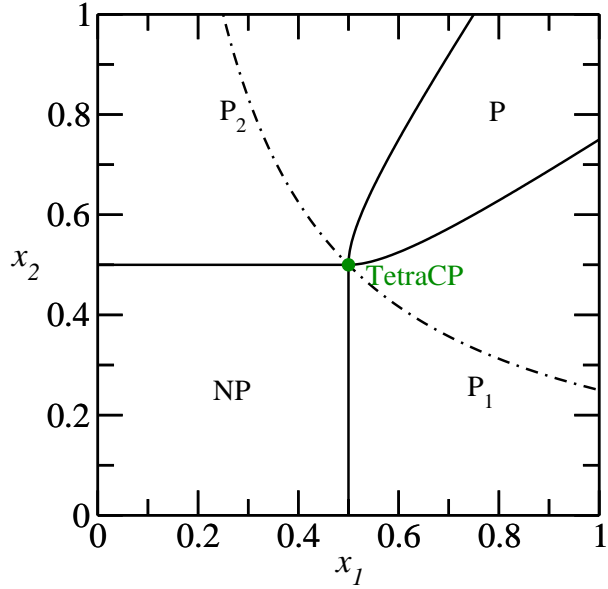}
\caption{(Color online) Phase diagram for $\kappa=1/4$. The  full (black) 
lines are continuous transitions. The green dot is a tetracritical point. 
In this and the following diagrams, on the dot-dashed  (black) curve the 
interactions between the polymer bonds vanishes ($\chi=1$).}
\label{q3k25}
\end{figure}

\begin{figure}[ht]
\centering
\includegraphics[width=8cm]{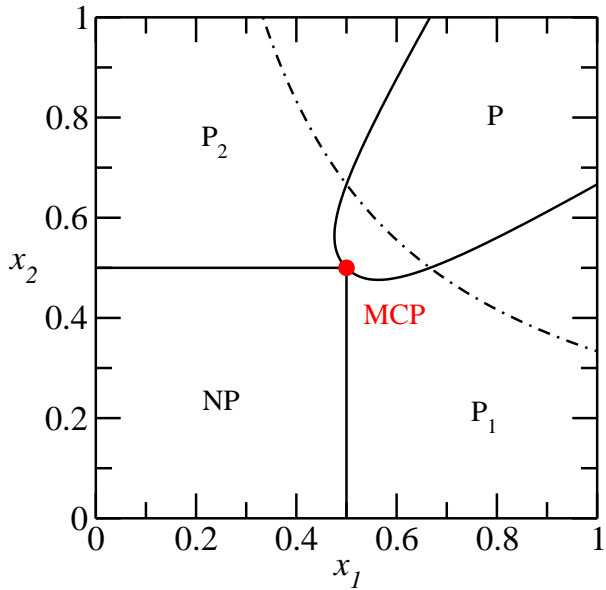}
\caption{(Color online) Phase diagram for $\kappa=1/3$. All transitions 
are continuous. The point where the four critical lines meet is the 
endpoint of a line of tetracritical points for $\kappa<1/3$ and four 
other lines, two of them tricritical and two of critical endpoints, 
for $\kappa>1/3$, thus being a multicritical point of higher order. The 
dot-dashed curve corresponds to $\chi=1$.}
\label{q3k033}
\end{figure}

\begin{figure}[ht]
\centering
\includegraphics[width=8cm]{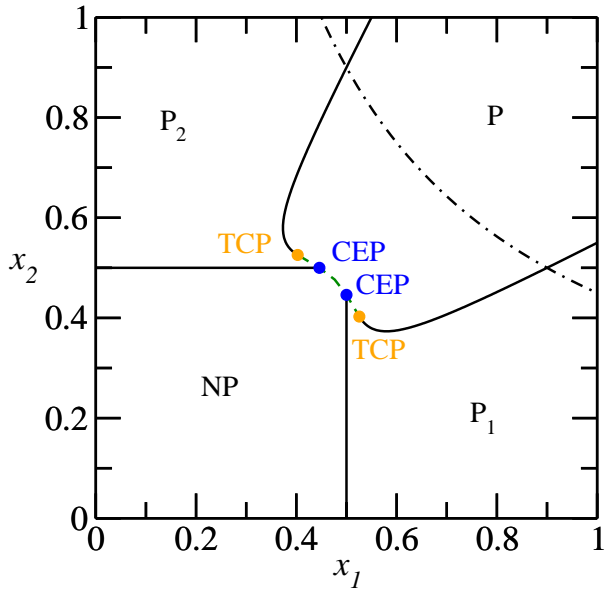}
\caption{(Color online) Phase diagram for $\kappa=9/20$. Two pairs of 
tricritical points (orange) and critical endpoints (blue) appear.The 
dot-dashed curve corresponds to $\chi=1$.}
\label{q3k045}
\end{figure}

\begin{figure}[ht]
\centering
\includegraphics[width=8cm]{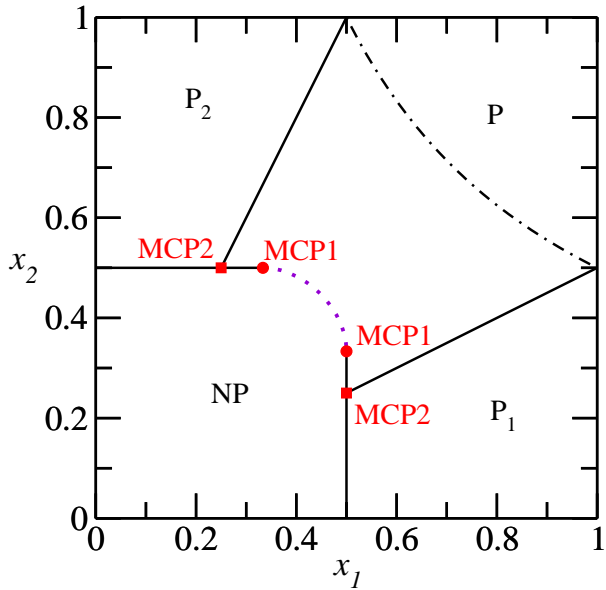}
\caption{(Color online) Phase diagram for $\kappa=1/2$. The dotted  
(violet) curve is a line of bicritical 
points, and is limited by two multicritical points (red dots). Between 
these multicritical points and the multicritical points represented by squares, 
the critical lines $P_2-P$ and $P-NP$, as well as $P_1-P$ and $P-NP$ 
coalesce, so that
on this line four critical surfaces meet, thus characterizing it as a 
tetracritical line (green).The 
dot-dashed curve corresponds to $\chi=1$.}
\label{q3k5}
\end{figure}

\begin{figure}[ht]
\centering
\includegraphics[width=8cm]{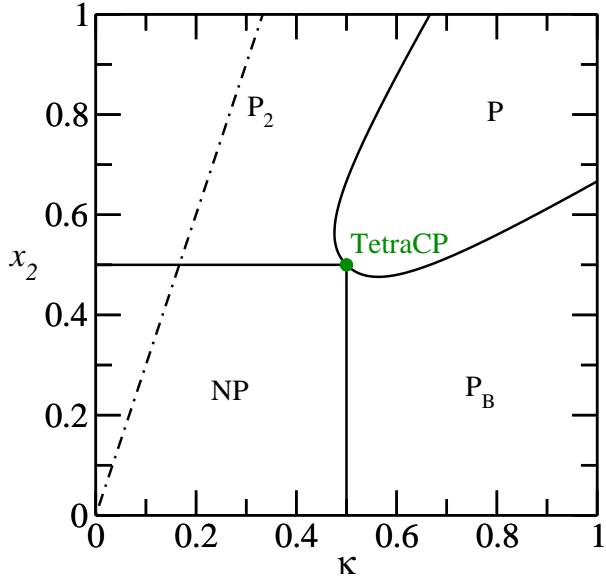}
\caption{(Color online) Phase diagram in the variables 
$(\kappa,x_2)$ for $x_1=1/3$. Four critical lines meet at a tetracritical 
point, represented by a circle (green).}
\label{q3kx2}
\end{figure}

\begin{figure}[ht]
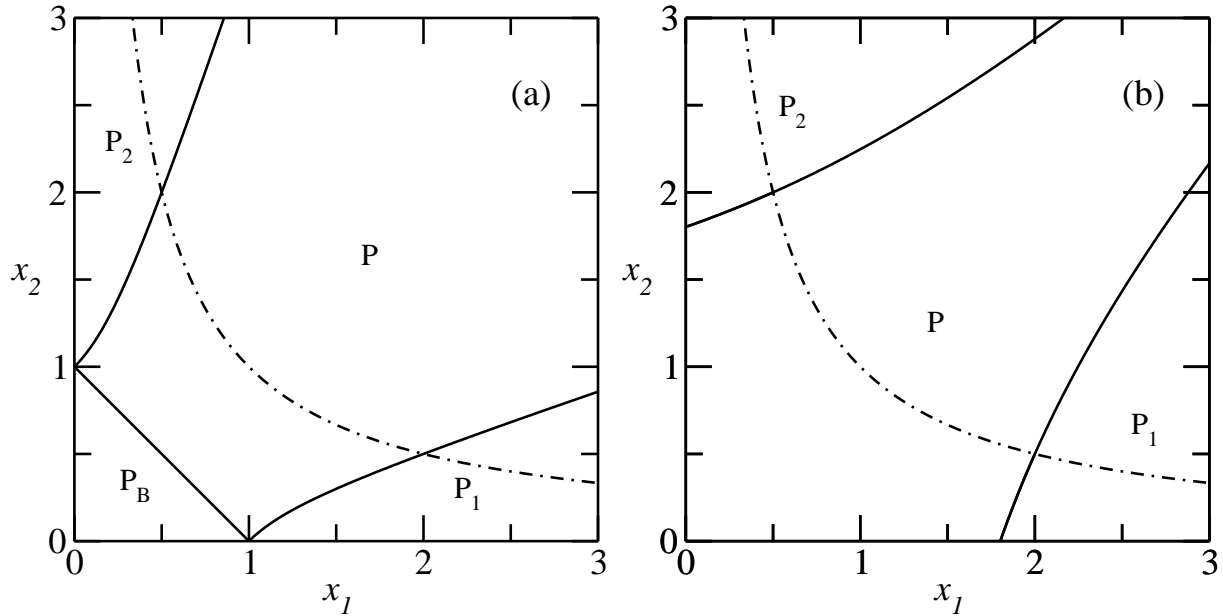

\centering
\includegraphics[width=8cm]{f9a.eps}
\includegraphics[width=8cm]{f9b.eps}
\caption{ Phase diagram for $\kappa=1$. The four polymerized
phases present in this diagram are separated by second order lines. The 
dot-dashed curve corresponds to $\chi=1$. The graph (a) is for the
model of interacting bonds, while the graph (b) is for the model with 
interaction between
 monomers.}
\label{q3k1}
\end{figure}

Since a rather large number of physical fixed points is found, the
model has a rich phase diagram, with different multicritical lines. 
In figures \ref{q3k0}-\ref{q3k1} we show six $(x_1,x_2)$ cuts for
$\kappa=0,\, 1/4,\, 1/3,\, 9/20,\,
1/2,\, 1$. These values of $\kappa$ were chosen  
in order  to illustrate the main features of the phase diagram.
The second-order lines and multicritical points were obtained analytically,
and the first-order lines were found numerically.

The first diagram, for $\kappa=0$, is depicted in figure
\ref{q3k0}. It actually corresponds to infinite {\em repulsive}
interactions between polymer bonds in corresponding edges of both
lattices. Although it should be said that this limiting case is
probably quite far away from real polymeric systems, the phase
diagram exhibits interesting features which lead us to shortly discuss
it here. If a polymerized phase exists in one of the
lattices, its presence inhibits the polymerization transition on the
other lattice, since it reduces the effective coordination number at
this other lattice. This effect is apparent in the diagram, the
transition to the phase $P_2$, at constant $x_1>1/(q-1)$ happens at
$x_2=x_1>1/(q-1)$ and is discontinuous, since when the free  energy of
the $P_1$ phase becomes smaller than the one of the $P_2$ phase, $x_1$
is larger than its critical value $1/(q-1)$. The coexistence line of the
two polymerized phases meets the two critical polymerization lines at a
point which is a bicritical point (BCP), located at
$x_1=x_2=1/(q-1)$. It is worth mentioning that the features of the
transition lines incident at the BCP are the ones expected for a
mean-field approximation as the one we are doing here: the classical
crossover exponent is $\phi=1$, so that the critical lines are linear
functions close to the BCP \cite{ba75}. In general, one finds $\phi>0$
and therefore the critical lines meet the coexistence line
tangentially, as may be found, for instance, from $\epsilon=4-d$
expansions for the $n$-vector model when $n>0$ up to order
$\epsilon^4$ \cite{k81}. However, all these coefficients vanish for $n
\to 0$, so that it is possible that even below the upper critical
dimension linear incidence of the critical lines at the BCP is
observed. 

When $\kappa$ has a small positive value, the $P$ phase occupies a
region between the $P_1$ and $P_2$ phases, separated from them by
critical lines, as may be seen in figure
\ref{q3k25}. With this intermediate phase present, the bicritical
point becomes a tetracritical point, similar to what is found in
anisotropic antiferromagnets \cite{ba75}. Again the remarks about the
crossover exponent $\phi$ and the incidence of the four critical lines
at the tetracritical point apply, and for a mean-field approximation
linear behavior is expected, in agreement with the results we
obtained. In this and the following diagrams, the dot-dashed curve
$x_1x_2=\kappa$ separates the regimes of repulsive ($\chi<1$) and
attractive ($\chi>1$) interactions between the polymers; the
tetracritical point is located on this curve. The region of 
attractive bond-bond interactions if situated below the curve. 
The two critical lines which limit the $P$ phase meet at a right
angle. Between $\kappa=1/4$ and $\kappa=1/3$, this angle is larger
than $\pi/2$, as may be seen in the phase diagram for $\kappa=1/3$
shown in figure \ref{q3k033}, where the angle is equal to $\pi$. 
A point which may be worth discussing is
that in this case a reentrant behavior is seen in the critical lines
which limit the phase $P$. This behavior is
actually a consequence of our choice of variables, since
$\kappa=x_1x_2\chi=const.$ implies that the Boltzmann factor of the
interactions $\chi$ is not constant if one of the activities
changes. In a similar diagram with constant $\chi$ the critical
curves are monotonous and no reentrance is seen.

As $\kappa$ exceeds $1/3$, a qualitative change of the phase diagram
happens. The tetracritical point splits into two pairs of tricritical
and critical endpoints, as may be seen in figure \ref{q3k045}. A first
order transition line now separates phase $P$ and $NP$, while part of
the transition between phase $P$ and the phases $P_1$ and $P_2$ becomes
discontinuous. As $\kappa$ grows, the tricritical points approach the
$NP-P_1$ and the $NP-P_2$ critical lines and the critical endpoints, so
that finally, at $\kappa=1/2$, the tricritical points meet the
critical surface which limit the $NP$ phase, as may be seen in figure
\ref{q3k5}. The multicritical points where the tricritical and critical 
endpoint lines meet are labeled MCP1 in the diagram. The dotted line 
which connects the two multicritical points is a line of bicritical 
points, since at this line two critical surfaces ($NP-PB$ and $PB-P$) and 
a coexistence surface ($NP-P$) meet. At this value of $\kappa$, the critical
lines between polymerized phases change from concave to convex, so
that they are linear.  The $NP$ phase is still stable in a region of the
phase diagram close to the origin, but its stability is quadratic, so
that the region covered by it in this phase diagram is actually a
critical surface separating it from the bilayer polymerized phase
$P_B$, which becomes stable for $\kappa>1/2$. On the lines between 
the multicritical 
points (MCP1 and MCP2), four critical surfaces meet ($NP-P_2$, $P_2-P$, 
$P-P_B$ and $P_B-NP$ on the upper part of the diagram), so that on this line 
we have tetracritical transitions. This feature may be 
better appreciated in the diagram in the variables $\kappa$ and $x_2$, for
$x_1=1/3$, which cuts this line of tetracritical points and is shown
in figure \ref{q3kx2}.

As soon as $\kappa$ exceeds $1/2$ all transitions are continuous. The
phase diagrams are similar to the one shown in figure
\ref{q3k1}, for $\kappa=1$. Graph (a) corresponds to the model
with interacting bonds. All four polymerized phases are present, 
there are three critical lines separating the phase $P$ from another 
polymerized
phase. Since the stability limits of the phases $P_1$, $P_2$, and
$P_B$ are known (Eqs. (\ref{p1l}) and (\ref{pbl})), the phase diagram
may be 
obtained analytically. The critical lines which limit the phases
$P_1$ and $P_2$ start at $x_1=0$ and $x_2=0$ respectively, thus
$\kappa=1/2$ is a quite singular value regarding the behavior of these
lines. As expected, the region of the parameter space where the
$P_B$ phase is stable grows with increasing values of $\kappa$.

While figure \ref{q3k0} is the same for interactions between bonds
or monomers, for nonzero values of $\kappa$ different phase diagrams
are found for both models. As  an example, in figure \ref{q3k1} we also 
show the result for the model with
interaction between monomers (b). We have chosen to compare the thermodynamic
properties of both models for larger values of $\kappa$ since obviously
the differences between the results for both models are bigger. We notice
that the bilayer polymerized phase is never the phase with
lowest free energy  for $\kappa=1$ if
the interaction is between monomers, although this phase is the one with
lowest free energy in the region of low values of
$x_1$ and $x_2$, for $1/2<\kappa<1$. The larger region in the parameter 
space where the $P$ phase has lower free energy as the $P_B$ phase, when
compared to the model with interacting bonds, may be qualitatively understood 
as follows. 
In the model with bond interaction the only configurations
where the interaction energy is minimized contribute to the phase $P_B$, with
a weight $(x_1x_2\chi)^N$, for a segment of $N$ double
steps. If the interaction
is between monomers, besides these configurations, there may be others with
the same statistical weight which contribute to the free energy of the 
phase $P$, thus lowering the free energy of this phase. The free energy of
the phase $P_B$ is the same in both models.

\subsection{Identical Polymers}

\begin{figure}[ht]
\centering
\includegraphics[width=8cm]{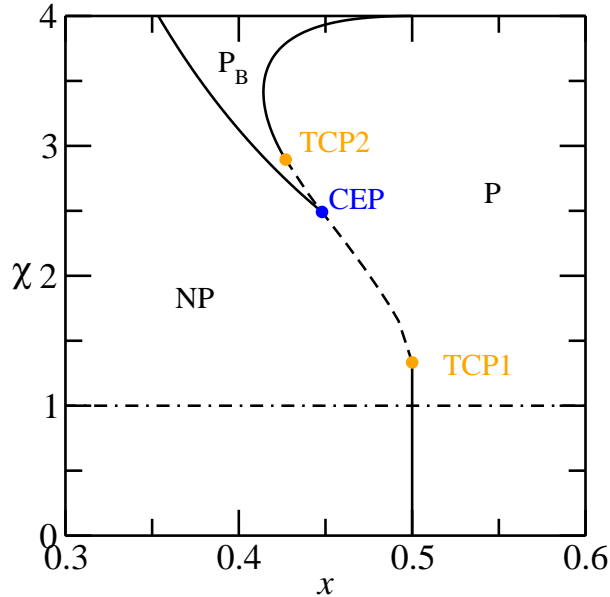}
\caption{(Color online) Phase diagram for $x_1=x_2=x$ and $q=3$. 
Two tricritical points (orange) and a critical end point (blue) 
shown.}
\label{chix}
\end{figure}

It is interesting to study the case  where both polymers are identical,
that is $x_1=x_2=x$. Besides the physical interest, this case is simpler
than the general one. Only the phases $NP$, $P$, and $P_B$ are present
in this subspace of the general parameter space discussed above. Since in
this case there is no particular advantage in using the composed variable 
$\kappa$, we use the Boltzmann factor $\chi$ instead. The phase 
diagram for $q=3$ if shown in figure \ref{chix}.
The line $x=1/(q-1)$ is a second-order line between phases $NP$ and
$P$ from $\chi=0$ up to a tricritical point localized at

\begin{equation}
x^{(TC1)}\,=\,\frac{1}{q-1} \;\;;\;\;\chi^{(TC1)}\,=\,\frac{q+1}{3}\;.
\end{equation}

For larger values of $\chi$, the $NP-P$ transition is discontinuous.
This part of the phase diagram is qualitatively identical to what 
if found for SASAW's, the tricritical point (TCP1) corresponds to the theta 
point where the SASAW's collapse. The $NP-P$ coexistence line ends
at a critical endpoint, whose location has been determined numerically,
since it involves the localization of the coexistence line.
The phases $P_B$ and $P$ coexist on a line which starts at the
critical endpoint and ends at the
a second  tricritical point (TCP2). Although it is possible to obtain an
analytical  expression for this  
tricritical  point as a function of $q$ using symbolic algebra programs, the 
resulting expressions are too long to be displayed here. In the particular case
 $q=3$, the tricritical point is given by
$(x_{TC2}=(3 \sqrt{5} -5)/4  , 
\chi_{TC2}=16(5 - 2 \sqrt{5})/(3\sqrt{5}-5)^2$). For values of 
$\chi > \chi_{TC2}$,
the $P_B-P$ line is a second order line given by
\begin{equation}
\chi\,=\,\frac{1}{x}\,\left(1+ (q-2) x \,\pm \, \sqrt{\frac{q-1}{q-2}\,
\left[(q-1) x (2+(q-2) x)-2 \right] }\right).
\end{equation}

An interesting
reentrant behavior is found in the critical line between the two 
polymerized phases: for $x$ constant and slightly below the tricritical 
value, as $\chi$ is increased one passes from the $P_B$ phase to 
the $P$ phase and them back to the $P_B$ phase. It may be understood that 
for high values of $\chi$ the bilayer polymerized phase is favored, since 
in it the internal energy is lower, as compared to the regular polymerized
phase, which is characterized by a larger entropy.

\section{Conclusion}
\label{conc}

As shown above, the phase diagram of polymers placed on a bilayer with 
first neighbor interactions between bonds in different layers displays a
variety of critical surfaces and multicritical lines and points, due to the
fact that one non-polymerized and four distinct polymerized phases appear. 
Although to our knowledge such a system has not yet be studied 
experimentally, it is possible that an experimental situation similar to
the model we study here may be found, with polymers confined to a region 
close to the interfaces between immiscible liquids, for example. Also, due to
the richness of the phase diagram of the model associated with 
the relative simplicity 
of its solution of the Bethe lattice, which even allows one to obtain many
features of the thermodynamic behavior analytically, we believe this model 
to have also some pedagogical interest for explaining multicritical points.

As discussed in the introduction, magnetic bilayers have been much studied
in the literature by a variety of theoretical methods, including pairs of
Bethe lattices similar to what was done above \cite{albayrak}, and, as 
was discussed in the introduction, polymer models may be mapped 
onto ferromagnetic $n$-vector
models in the limit $n \to 0$. This correspondence may be generalized along 
the lines proposed for equilibrium polymerization in poor solvents 
\cite{wp81b}, and
in a particular limit may map into an effective $n \to 0$-vector model 
with higher order spin interactions between spins in distinct layers, and
therefore it may be possible that similar phase diagrams to the ones found 
here may appear in related magnetic models, even for other values of $n$. Thus,
there exists a possibility that phase diagrams similar to the ones found here
may appear in appropriate magnetic bilayers.

We notice that in all phase diagrams, with the exception of the first
one shown in figure \ref{q3k0}, the most interesting features, such
as multicritical loci, appear in the region of 
parameters where the interaction between bonds is attractive ($\chi>1$),
and this corresponds to the effective models for polymers in poor 
solutions mentioned in the introduction. 

Although we did not present here in detail the results for the model
with interactions between monomers, in the particular case $\kappa=1$
where we compared both models it is already apparent that the results are
qualitatively different. This may be seen as rather surprising at first,
but, as mentioned above, happens for SASAW's in two-dimensional lattices 
\cite{mos01} and for Husimi lattices built with squares \cite{sms96}, where
a saturated polymerized phase is stable in a region of the parameter phase
for the model with interactions between bonds.
The attractive interactions between bonds and monomers both favor
more compact polymerized phases, but the details of these phases and
the phase diagram may be quite different for both models. 
The richness of the phase diagram of polymers on bilayers is quite impressive,
four distinct polymerized phases appear, and this gives rise to a variety
of phase transition and multicritical loci.

\acknowledgments
PS acknowledges CONICET, SECYT-UNC, and MinCyT C\'ordoba
for partial financial support of this project and the hospitality at
IF-UFF, as well as support by Secretar\'{\i}a de 
Pol\'{\i}ticas Universitarias under grant PPCP007/2012.
JFS acknowledges the hospitality
at FaMAF-UNC and financial support by CAPES, though project 
CAPES/Mercosul PPCP 007/2011, and by CNPq.

\end{document}